\documentclass[11pt]{article}

\usepackage{latexsym}

\def\a{\alpha}
\def\b{\beta}
\def\C{\raise2pt\hbox{\rm\large$\chi$}}
\def\c{\chi}
\def\d{\delta}
\def\e{\epsilon}
\def\g{\gamma}
\def\h{\eta}

\def\j{\psi}
\def\k{\kappa}
\def\l{\lambda}
\def\m{\mu}
\def\n{\nu}

\def\p{\pi}
\def\q{\theta}
\def\r{\rho}
\def\s{\sigma}

\def\x{\xi}

\def\D{\Delta}
\def\F{\Phi}

\def\vq{\vartheta}

\def\tb{\tilde{b}}

\DeclareFontFamily{OT1}{msb}{}{}
\DeclareFontShape{OT1}{msb}{m}{n}
 {  <5> <6> <7> <8> <9> <10> gen * msbm
      <10.95><12><14.4><17.28><20.74><24.88>msbm10}{}
\DeclareMathAlphabet{\bubble}{OT1}{msb}{m}{n}

\def\bR{{\bubble R}}
\def\bZ{{\bubble Z}}
\def\buC{{\bubble C}}

\def\ca{{\cal A}}

\def\cd{{\cal D}}

\def\cl{{\cal L}}

\def\cp{{\cal P}}

\def\cs{{\cal S}}

\def\det{{\rm det}}
\def\PCO{\cp}
\def\SFO{\cs}

\def\tPCO{\tilde{\cp}}

\def\half{{1\over 2}}
\def\pa{\partial}

\def\bo{{\raise-.5ex\hbox{\large$\Box$}}}

\def\ra{\rightarrow}

\def\Tr{{\rm Tr}}
\def\sdot{{\cdot}}

\def\ket#1{\left| #1\right\rangle}              
\def\VEV#1{\left\langle #1\right\rangle}        
\def\leftrightarrowfill{$\mathsurround=0pt \mathord\leftarrow \mkern-6mu
        \cleaders\hbox{$\mkern-2mu \mathord- \mkern-2mu$}\hfill
        \mkern-6mu \mathord\rightarrow$}
\def\dvec#1{\vbox{\ialign{##\crcr
        \leftrightarrowfill\crcr\noalign{\kern-1pt\nointerlineskip}
        $\hfil\displaystyle{#1}\hfil$\crcr}}}           
\def\dtt#1{{\buildrel {\hbox{\LARGE .}} \over {#1}}}     
\def\ad{\dtt{\a}}
\def\da{\dtt{\a}}
\def\bd{\dtt{\b}}
\def\db{\dtt{\b}}
\def\pd{\dtt{+}}
\def\md{\dtt{-}}

\def\fr#1#2{{\textstyle{#1\over\vphantom2\smash{\raise.20ex
        \hbox{$\scriptstyle{#2}$}}}}}                   

\def\beq{\begin{equation}}
\def\eeq{\end{equation}}
\def\beqx{\begin{displaymath}} 
\def\eeqx{\end{displaymath}}
\def\beql{\arraycolsep .1em \begin{eqnarray}}
\def\eeql{\end{eqnarray}}
\def\zeile{\nonumber\\[.5ex] }
\def\gl#1{(\ref{#1})}

\def\theequation{\thesection.\arabic{equation}}

\catcode`@=11
\@addtoreset{equation}{section}
\@addtoreset{equation}{subsection}
\def\theequation{\ifnum\value{section}=0 \arabic{equation}\ignorespaces
\else \ifnum\value{section}=-1 A.\arabic{equation}\ignorespaces
\else \ifnum\value{subsection}=0 \thesection.\arabic{equation}\ignorespaces
\else \thesection.\arabic{subsection}.\arabic{equation}\ignorespaces
                           \fi
                      \fi
                 \fi}
\catcode`@=12

\parskip=\medskipamount

 \def\Amm{A^{--} } \def\Dmm{\cd^{--} }
\def\dpm{\pa^{\pm\pm}}  \def\Dpm{\cd^{\pm\pm}}
\def\nab{\nabla}
 
\def\sM{${\widehat {\cal M}}^+$} \def\hM{${\widehat {\cal M}}$}
\def\dim#1{dimension $#1$ }
\def\0#1{{\stackrel{\circ}{#1}}} 
\def\der#1{{\partial \over \partial #1}}

\def\N#1{$N{=}#1$}
\def\square{\kern1pt\vbox
            {\hrule height 0.6pt\hbox{\vrule width 0.6pt\hskip 3pt
 \vbox{\vskip 6pt}\hskip 3pt\vrule width 0.6pt}\hrule height 0.6pt}\kern1pt}

\def\hs{harmonic space} 
\def\ji{Jacobi identity}

\def\sd{self-dual}
\def\sdy{self-duality}
\def\sp{super-Poincar\'e algebra}
\def\ym{Yang-Mills}

\def\be{\begin{equation}}
\def\ee{\end{equation}}
\def\la#1{\label{#1}}  
\def\arr{\begin{array}{rll}}
\def\ea{\end{array}}
\def\bea{\begin{eqnarray}}
\def\eea{\end{eqnarray}}

\def\eb{{\bar \eta}} 
\def\f{\varphi}

\begin{document}


\begin{titlepage}

\noindent
hep-th/9712043 
\hfill ITP--UH--31/97 \\

\vskip 1.0cm

\begin{center}

{\Large\bf EXTENDED SELF-DUAL YANG-MILLS}\\

\medskip

{\Large\bf FROM THE N=2 STRING~$^*$}\\

\vskip 1.5cm

{\large Chandrashekar Devchand}

{\it Max-Planck-Institut f\"ur Mathematik in den Naturwissenschaften}\\
{\it Inselstra\ss e 22-26,  04103  Leipzig, Germany}\\
{E-mail: devchand@mis.mpg.de}\\

\vskip 0.7cm

{\large Olaf Lechtenfeld}

{\it Institut f\"ur Theoretische Physik, Universit\"at Hannover}\\
{\it Appelstra\ss{}e 2, 30167 Hannover, Germany}\\
{http://www.itp.uni-hannover.de/\~{}lechtenf/}\\

\vskip 1.5cm
\textwidth 6truein
{\bf Abstract}
\end{center}
\begin{quote}
We show that the physical degrees of freedom of the critical open 
string with \N2 superconformal symmetry on the worldsheet are described 
by a self-dual Yang-Mills field on a {\it hyperspace\/} parametrised 
by the coordinates of the target space $\bR^{2,2}$ together with a 
{\it commuting\/} chiral spinor. A prepotential for the self-dual 
connection in the hyperspace generates the infinite tower of physical 
fields corresponding to the inequivalent pictures or spinor 
ghost vacua of this string. An action is presented for this tower, which  
describes consistent interactions amongst fields of arbitrarily high spin. 
An interesting truncation to a theory of five fields is seen to have no 
graphs of two or more loops.

\end{quote}

\vfill

\textwidth 6.5truein
\hrule width 5.cm
\vskip.1in

{\small \noindent ${}^{*\ {}}$
supported in part by the `Deutsche Forschungsgemeinschaft';
grant LE-838/5-1}

\end{titlepage}

\hfuzz=10pt


\section{Introduction}

The \N2 string\footnote{
For a review of the subject until 1992, see ref.~\cite{marcus}.}
has many unique features descending from various remarkable 
properties of the $(1{+}1)$-dimensional \N2 superconformal algebra, 
the gauge symmetry on the worldsheet. In particular, these 
superconformal gauge symmetries kill all oscillatory modes of the string, 
yielding a peculiar string without any massive modes. 
The \N2 worldsheet supersymmetry is the maximal one for which the 
perturbative (e.g. BRST) quantisation yields a positive critical dimension. 
This turns out to be four~\cite{AL}. However, the target space coordinates 
carry a complex structure which implies a euclidean $(4,0)$ or kleinian 
$(2,2)$ signature. 
In the target space, the (naive) single degree of freedom corresponds to 
a scalar field, which in the open (resp. closed) sector is the dynamical 
degree of freedom of a self-dual gauge (resp. graviton) field~\cite{OVold}.  
The absence of physical oscillator excitations of the string also implies 
that all scattering amplitudes beyond the three-point function 
vanish~\cite{OVold,hippmann}. 
The relation to self-duality provides this string with its particularly 
rich geometric structure, and we shall describe a further novel feature 
in this paper. We restrict ourselves to the open string sector, though 
analogous arguments hold for the closed sector as well; 
and we adopt the more interesting case of indefinite signature metric
and specialise to a flat target space $\bR^{2,2}$~\cite{klein}.

Perturbative open string Hilbert spaces are defined using the
{\it relative\/} cohomology of the BRST operator $Q_{BRST}$
on the Fock space of open string excitations.
In other words, {\it physical\/} open string states correspond to
elements of the coset $im\ Q_{BRST}/ker\ Q_{BRST}$, after imposing
the subsidiary conditions $b_0=0=\tb_0$.
Here, $b_0$ and $\tb_0$ are the anticommuting antighost zero modes of
the open \N2 string;
and we do {\it not\/} demand further restrictions 
involving the spinor antighost zero modes $\b^\pm_0$.
Chiral bosonisation of the spinor ghosts~\cite{FMS} enlarges the open \N2 
string Fock space. It is then no longer graded by only the mass level and 
the {\it total ghost number\/} $u\in\bZ$, but acquires {\it two\/} 
additional gradings: the {\it picture numbers\/} $\p_\pm$ labelling 
inequivalent spinor ghost vacua called {\it pictures\/}. These are related 
by {\it spectral flow\/} ($\SFO$) and {\it picture-raising\/} ($\PCO^\pm$) 
transformations which commute (up to BRST-exact terms) with one another 
as well as with $Q_{BRST}$~\cite{BKL,pope,KL,buckow,BL1,BL2,LS}.
These maps, however, do not afford a complete equivalence between the 
{\it relative\/} BRST cohomologies in the different pictures\footnote{
A local picture-lowering operator inverting $\PCO^\pm$ does not exist in the 
\N2 string~\cite{BKL,pope}. It occurs in the \N1 string, where, because it 
does not commute with $b_0$, it guarantees picture-equivalence~\cite{HMM} 
only for the {\it absolute\/} BRST cohomology on the unrestricted Fock space 
obtained by dropping the subsidiary conditions.}, 
although on states with non-zero momentum it is possible to establish 
an equivalence~\cite{berk3,JL}.
Further analysis proves that each picture contains exactly {\it one\/}
(absolute as well as relative) BRST cohomology class 
for a given non-zero lightlike momentum~\cite{JL}.

The physical state in each picture, however, changes under target space
$SO(2,2)$ action in a fashion characteristic of the particular picture. 
Specifically, it transforms as a highest weight state of a spin~$j$ 
representation of the non-manifest $SL(2,\bR)$ subgroup of $SO(2,2)$.  
The spin $j=0,\fr12,1,\fr32,\ldots $ may be used as a convenient picture 
label instead of the {\it total picture number\/} 
$\p_+{+}\p_-\in\bZ$~\cite{BL1,BL2,LS}. A fully $SO(2,2)$-covariant 
formulation of the \N2 string cannot be expected to identify 
unequal-multiplicity-states in different pictures. This reinforces 
the notion that the pictures include physically distinct states. 
We adopt the attitude that the total picture number is a physical 
quantum number for \N2 string excitations.
In contrast, there is no physics in the difference $\p_+{-}\p_-$.
We stress that the situation is completely different from the \N1 case.
There, only {\it one\/} picture grading by $\p\in\fr12\bZ$ appears. 
Most importantly, physical states in pictures with $\p$ differing by an 
integer have identical target space transformation properties and are 
therefore rightly identified, yielding only a single NS ($\p\in\bZ$) 
and a single R ($\p\in\bZ{+}\fr12$) sector.

\N2 supermoduli transformations include twists of the $U(1)$ gauge bundle 
generated by {\it spectral flow\/}~\cite{SS} 
relating the one-parameter family of twists of the fermionic boundary 
conditions interpolating between periodic (NS) and antiperiodic (R).
It also changes $\p_+{-}\p_-$ but not the total picture number.
Since the quantum theory involves a functional integral over the 
supermoduli, there is no physical distinction between sectors having 
different boundary conditions, and all sectors are equivalent to the 
NS sector ($\p_+{-}\p_-\in\bZ$). This property implies the absence of 
target space fermions, and therefore this string fails to reflect its 
worldsheet supersymmetry in a target space supersymmetry.  
However, as we shall demonstrate, it realises an alternative extension of 
the Poincar\'e algebra, obtainable from the \N1 \sp\ by changing the 
statistics of the Grassmann-odd (fermionic) generators. The algebra thus 
obtained is a {\it Lie\/} (rather than super) extension of the Poincar\'e
algebra by Grassmann-even (bosonic) spin $\half$ generators.  

This extended Lie algebra
is a genuine symmetry algebra on the space of physical states,
with picture-raising being interpreted as an {\it even\/} variant of a
supersymmetry transformation. The earlier viewpoint~\cite{OVold} of a 
one-dimensional physical state space is thus revealed to be a (consistent) 
truncation of an infinite tower of physical states of increasing spin. 
We present an effective action for this infinite tower of fields
and demonstrate it to be a component version of an extension of self-dual 
Yang-Mills to a {\it hyperspace} with standard vectorial coordinates 
$x^{\a\da}$ supplemented by an even (commuting) chiral spinor $\eta^\a$.
A prepotential for the self-dual connection in the hyperspace is shown
to generate the entire tower of physical fields.

The plan of this paper is as follows. In section 2 we review some relevant
features of \N2 string theory, describing in particular the infinite set of
superconformally inequivalent physical states. In section 3, by considering 
the tree-level scattering amplitudes for these physical states, we deduce an
effective action, $S_\infty$, for the corresponding infinite set of target 
space fields. This action is seen to consistently truncate to a two-field 
action previously considered in \cite{CS}, as well as a novel five-field 
action. In section four we rederive $S_\infty$ from a consideration of
generalised self-dual Yang-Mills on an even-spinorial extension of $\bR^{2,2}$,
mimicking the construction of superspace using spinorial coordinates of
the `wrong' statistics. On a chiral subspace of this hyperspace, the 
well known Leznov functional \cite{leznov,parkes} is then seen to be a
hyperspace-covariant version of $S_\infty$. Finally, we present an 
$SO(2,2)$-invariant action for the five-field model, reminiscent of 
the action for $N{=}4$ {\it supersymmetric\/} self-dual 
Yang-Mills~\cite{siegel}.

\vfill\break


\section{N=2 Open String Worldsheets}

Strings with two world-sheet supersymmetries in the NSR formulation
are built from an $N{=}2$ world-sheet supergravity multiplet containing
the metric $h_{mn}$, a $U(1)$ gauge field $a_m$, and two charged Majorana
gravitini $\c^\pm_m$.
Quantum consistency demands $c_{\rm matter}=6$,
corresponding to a target space of real dimension four.
Analysis of $N{=}(2,2)$ non-linear $\s$-models produces three
distinct possibilities for worldsheet matter fields~\cite{sevrin}:\\
(a) two chiral superfields\\
(b) one chiral and one twisted-chiral superfield\\
(c) one semi-chiral superfield.\\
Since case (c) has not yet been much studied, and case (b) leads to
free strings only~\cite{hull}, we will concentrate on the standard case (a),
in which the component content of the worldsheet matter is given by the
four string coordinates $X^\m$ and their $U(1)$ charged NSR partners $\j^\m$.

$N{=}2$ worldsheet supersymmetry implies a target space complex structure,
so the (real) spacetime metric must have signature $(2,2)$
if we require light-like directions.
Specialising to a flat target space, $\bR^{2,2}$, we write
\beq
X^{\a\ad}\ =\ \s_\mu^{\a\ad} X^\mu\ =\
\left(\begin{array}{cc}
X^0{+}X^3 & X^1{+}X^2 \\ X^1{-}X^2 & X^0{-}X^3
\end{array}\right) \quad,\qquad
\a\in\{+,-\} \quad \ad\in\{\pd,\md\} \;,
\eeq
with a set of chiral gamma matrices $\s_\mu$, $\mu=0,1,2,3$,
appropriate for a spacetime metric $\h_{\mu\nu}={\rm diag}(-+-+)$.
We use the van der Waerden index notation,
splitting $SO(2,2)$ vector indices $\mu$ into two $SL(2,\bR)$
spinor indices, $\a$ and~$\ad$, which are raised and lowered using the 
$SL(2,\bR)$-invariant skew-symmetric tensor, e.g.,
$\k\cdot\l\ =\ \k^\a\l_\a\ =\ \e_{\a\b}\k^\a\l^\b\ =\ \e^{\a\b}\k_\b\l_\a\ $;
and vectors have an $SL(2,\bR)\times SL(2,\bR)'$ invariant length-squared
\beq
\h_{\mu\nu}\ X^\mu X^\nu\
=\ -\fr12\,\e_{\a\b}\,\e_{\ad\bd}\ X^{\a\ad}\,X^{\b\bd}\
=\ -\det\,X^{\a\ad} \quad.
\eeq
The NSR formulation~\cite{brink} requires a specific choice of 
complex structure, $\bR^{2,2}\to\buC^{1,1}$, thus breaking this $SO(2,2)$ 
transformation group to a subgroup leaving the complex structure invariant,
namely, 
\beq
SL(2,\bR)\times SL(2,\bR)'\ \to\ GL(1,\bR)\times SL(2,\bR)' \quad.\la{subgp} 
\eeq
The residual $SL(2,\bR)$ transformations change the complex structure. 
Indeed, they generate the complex structure moduli space, 
$SL(2,\bR)/GL(1,\bR)$, where $GL(1,\bR)$ is a parabolic subgroup.
We choose a representation for the $sl(2,\bR)$ algebra 
$\{ L_{\pm\pm},L_{+-}\}$, with $L_{+-}$ diagonalised as one of the 
{\it boost\/} generators, having eigenvalues $m$  called {\it boost charges\/}.
The rotation and the second boost are generated by linear combinations
of the nilpotent $L_{++}$ and $L_{--}$. Then, following~\cite{LS},
we may choose the unbroken $gl(1,\bR)$ generator to be $L_{++}$.

As mentioned in the Introduction, the most elegant way to classify 
physical open string states
is by solving the relative cohomology of the BRST operator~$Q_{BRST}$.
This yields a spectrum of only massless states distinguished by
the total ghost number $u\in\bZ$ and a pair of picture charges
$(\p_+,\p_-)$ labelling inequivalent spinor ghost vacua.
For physical states these three quantum numbers are related by
$u=\p_++\p_-+1$, so that in every picture there exists exactly one 
cohomology class for a given non-zero lightlike momentum~\cite{JL}.
It is often convenient to use the sum and difference,
\beq\begin{array}{rcl}
\p\ &\equiv\ & \p_++\p_-\ \in\bZ\qquad{\rm total\ picture}\zeile
\D\ &\equiv\ & \p_+-\p_-\ \in\bR\qquad{\rm picture\ twist}\quad,
\end{array}\eeq
and denote states like $\ket{\p;\ldots}_\D$.
Then, in the picture $(-1,-1)$, the BRST analysis~\cite{bien}
leads to a single physical state~\cite{OVold} with $u{=}-1$, namely,
\beq
\ket{-2;k}_0\ =\ V(k)\ket{0;0}_0
\eeq
where $V(k)$ represents the string vertex operator for the
center-of-mass mode with lightlike momentum~$k^\m$.
Nothing else appears. In particular, the massive states one would
naively expect of a string do not materialise.
Since we are dealing with open strings, a Chan-Paton adjoint gauge index,
which we generally suppress, is to be assumed.
It should be noted that null vectors like~$k^\m$ factorise into spinors,
\beq
\h_{\m\n}\ k^\m k^\n\ =\ 0 \qquad\Leftrightarrow\qquad
k^{\a\ad}\ =\ \k^\a\ \k^\ad \quad, \la{factor}
\eeq
which simplifies the massless dynamics.

The picture twist may be changed continuously, $\D\to\D{+}2\r$, by 
applying the spectral-flow operator $\SFO(\r)$, $\r\in\bR$.\footnote{
The argument of $\SFO$ is not compact. Writing $\r=c+\vq/2\p$,
$\vq$ is the angle of spectral flow and $c$ is the change
in $U(1)$ instanton number~\cite{KL,buckow}.}
The $\SFO(\r)$ form an abelian algebra with zero ghost number, and they 
commute with $Q_{BRST}$, $\tPCO^\pm$, and $\tb_0$ but not with $b_0$.
Nevertheless, since the $N{=}2$ string integrates over the parameter of 
spectral flow, it identifies states of differing picture twist~$\D$
for fixed~$\p$. For convenience, we shall use the $\D{=}0$ representative 
and drop the $\D$ label from now on.

Can we also identify physical states having different $\p$?
As in the \N1 case, picture-raising commutes (up to BRST-exact 
terms) with both $Q_{BRST}$ and the antighost zero-modes $b_0$ and $\tb_0$. 
As for the \N1 string, picture-lowering with the same properties can be defined 
on states with non-zero momentum~\cite{berk3}.
Consequently, for $k\sdot k{=}0$ but $k{\neq}0$, 
an equivalence relation exists between the single relative cohomology classes 
appearing in any two pictures. From the worldsheet point of view,
it may therefore seem reasonable to conjecture picture equivalence 
to identify all physical states, yielding a single massless scalar field's 
worth of physical degree of freedom~\cite{OVold}.
However, things are not so simple~\cite{BL1,BL2,LS}, for there exist 
{\it two\/} such picture-raising operators,
$\tPCO^\a = \tPCO^\pm$, as components of an $SL(2,\bR)$ {\it spinor\/}.
They do not twist the pictures ($\D{=}0$)
and commute modulo BRST exact terms.\footnote{
These are not the naive picture-raising operators
but contain a spectral flow factor.}
Of course, the difference $\tPCO^+{-}\tPCO^-$ is BRST exact, but in 
a non-local way (involving division by momentum components).
In the following, we shall argue against the identification 
$\tPCO^+{\cong}\tPCO^-$ in view of target space properties.

Clearly, iterated picture-raising on $\ket{-2;k}$ creates 
states of $\p{=}-2{+}2j$ thus:
\beq
\tPCO^{(\a_1}\tPCO^{\a_2}\ldots\tPCO^{\a_{2j})} \ket{-2;k}\ =\
\ket{-2{+}2j; (\a_1\a_2\ldots\a_{2j}),\ k} \ ,\quad
j=0,\fr12,1,\fr32,\ldots \la{sl2states}
\eeq
where the extra label $(\a_1\a_2\ldots\a_{2j})$ denotes the tensorial
transformation property under $SL(2,\bR)$; all states being singlets 
with respect to $SL(2,\bR)'$. We emphasise that states of both integer and 
half-integer spin $j$ have the {\it same\/} (viz. bosonic) statistics. The
appearance of a $(2j{+}1)$-dimensional tensor representation would 
seem to contradict the earlier statement of unit multiplicity
for each picture~\cite{JL}. However,
as we have already mentioned, any specific choice of complex structure 
breaks $SO(2,2)$ as in \gl{subgp}; and the physical state in a 
given picture is indeed a singlet under the manifest transformation group
$GL(1,\bR)\times SL(2,\bR)'$. On the other hand, the  
transformations in the coset $SL(2,\bR)/GL(1,\bR)$ not only change 
the complex structure but also transform the physical states according 
to spin~$j$ $SL(2,\bR)$-representations, where $j$ depends on the 
picture~\cite{BL1,BL2,LS}. The string thus 
supports $SL(2,\bR)$ multiplets of physical states of {\it any\/} spin, 
but the specific choice of complex structure is tantamount to a
projection to a highest weight state of the $SL(2,\bR)$ multiplet. 
The components of the $SL(2,\bR)$ multiplets in \gl{sl2states} 
are therefore related to each other by changes of the complex structure. 
In fact, they can all be considered to be different components 
of a single physical state, the linear combination 
\beq
\ket{-2{+}2j; e,\q, k}\ :=\  
\sum_{(\a_1\a_2\ldots\a_{2j})} v_{\a_1}v_{\a_2}\ldots v_{\a_{2j}}\ 
\ket{-2{+}2j; (\a_1\a_2\ldots\a_{2j}),\ k}\quad, \la{comb}
\eeq
where the components of the spinor $v_\a$ parametrise the two 
dimensional parabolic coset of complex structures, $SL(2,\bR)/ GL(1,\bR)$. 
Conversely, it turns out that
\beq
\ket{-2{+}2j; (\a_1\a_2\ldots\a_{2j}),\ k}\ \propto\
\k^{\a_1}\k^{\a_2}\ldots\k^{\a_{2j}}\ \ket{-2{+}2j; e,\q, k}\quad,
\eeq
i.e. the $SL(2,\bR)$ charges are carried exclusively by momentum spinors
appearing in the normalisation of the state. We remark that the 
proportionality factor becomes singular when $v\sdot\k{=}0$.

On the worldsheet 
the parameters of the space of complex structures correspond to
the (open) string coupling~$e$ and the worldsheet instanton angle~$\q$.
For fixed values of $e$ and~$\q$, there exists therefore one physical
state in each picture.
Following~\cite{BL1,LS} we may put
\beq
\biggl( \begin{array}{c} v_+ \\ v_- \end{array} \biggr) \ =\ \sqrt{e}\
\biggl( \begin{array}{c} \cos\fr{\q}{2} \\ \sin\fr{\q}{2} \end{array}
\biggr) \quad. \la{vwahl}
\eeq
This amounts to prescribing non-trivial transformation behaviour
for the string coupling ``constants'': 
$e$ is a boost velocity, and $\q$ a rotation angle!
If we further choose a $\q{=}0$ complex structure, 
only the highest $SL(2,\bR)$ weights (all $\a_i{=}+$) survive. 
In particular $\tPCO^-$ disappears.

What about pictures with $\p{<}-2$?
Contrary to previous belief~\cite{LS}, the BRST cohomology is not trivial 
there, but again yields a single massless state~\cite{JL} in each picture.
Indeed, for generic light-like momentum, with the factorisation~\gl{factor} 
denoted $k=\k \dtt\k$, 
the physical Fock space has a non-degenerate scalar product,
\beq
\VEV{-k, \q, e; {-}2{-}2j\ |\ {-}2{+}2j; e, \q, k}\ \propto\ (v\cdot\k)^{2j}
\qquad{\rm for}\quad j\ge0
\eeq
where the conjugate states in pictures $\p=-2{-}2j$
are constructed in a fashion analogous to the $\p{=}-2{+}2j$ states,
but with conjugate spinors $\bar{v}$ and $\bar{\k}$ satisfying
\beq
\bar{v}\cdot v\ =\ 1 \qquad{\rm and}\qquad \bar{\k}\cdot\k\ =\ 1 \quad.
\eeq
In the absence of a standard picture-lowering operator, we do not know
a direct way of obtaining physical $\p{<}-2$ states from $\ket{-2;k}$.
Since the states $\ket{-2{+}2j;(\a_1\a_2\ldots\a_{2j}),\ k}$, for fixed $j,k$, 
are all proportional to one another, the metric in the spin~$j$ representation
space has rank one. Nevertheless, the conjugate states
\beq
\ket{-2{-}2j; (\a_1\a_2\ldots\a_{2j}),\ k} \quad, \qquad j>0 \quad,
\eeq
form an $SL(2,\bR)$ representation of the same spin, with highest and
lowest weights interchanged.  Hence, 
only $\a_i{=}-$ survives in the $\ket{-2{-}2j; e, \q{=}0, k}$ representative.
It is convenient to label {\it all\/} the states using $\p=-2{+}2j$, 
allowing for negative spin~$j$. 
The pattern which emerges is symmetric around $\p{=}-2$.

Having determined the physical states, string amplitudes are computed
by choosing a set of physical vectors as external states and integrating
their product with the string measure over the \N2 supermoduli. 
This simplifies at tree-level to a sum over different world-sheet topologies 
in the form of $U(1)$ instanton sectors classified by the
first Chern number~$c\in\bZ$ of the principal $U(1)$ gauge bundle.
Inspection of the
string path integral shows that only the range $|c|\leq J:=n{-}2$
contributes to the $n$-point scattering amplitude~\cite{berk1}.
Using only the representatives $\ket{-2{+}2j; e,\q, k}$ in \gl{comb}
as external states, with the selection rule
\beq
\sum_{s=1}^{n} j_s \ =\ J \quad,
\eeq
the instanton sum is automatically generated,
with the correct weights\footnote{Since $J>0$, all $j_s$ can be chosen 
to be non-negative and so no $\bar{v}$'s need appear.}
\beq
v_{\a_1} v_{\a_2} \ldots v_{\a_{2J}} \ =\ v_+^{J+c}\ v_-^{J-c}\ =\
e^J\ \cos^{J+c}\fr{\q}{2}\ \sin^{J-c}\fr{\q}{2}
\eeq
multiplying the individual contributions carrying boost charges
$m=c=-J,\ldots,+J$~\cite{BL1}.
All terms for a fixed value of~$c$ are identical, since one may picture-change
the open spinor indices freely from one external state to another.
It has been shown that all tree-level amplitudes vanish~\cite{hippmann},
except for the two- and three-point functions.
The latter is given by
\beq
A(k_1,k_2,k_3; e,\q)\ =\
f^{a_1 a_2 a_3}\ v_\a k^{\a\ad}\ v_\b k^{\b\bd}\ \e_{\ad\bd}\ =\
f^{a_1 a_2 a_3}\ v\sdot\k_1\ v\sdot\k_2\ \dtt{\k}_1\sdot\dtt{\k}_2
\eeq
where $f^{a_1 a_2 a_3}$ are the (antisymmetric) structure constants of the 
gauge algebra, $a_i$ are Chan-Paton indices and the momenta 
$k_i = {\k}_i \dtt{\k}_i, (i=1,2,3)$  satisfy
\beq
k_1+k_2+k_3=0 \qquad{\rm and}\qquad
\k_i\cdot\k_j=0 \quad.
\eeq
This amplitude is totally symmetric under interchange of the three legs.
For the choice $\q{=}0$, it reduces to
\beq
A(k_1,k_2,k_3; e,\q{=}0)\ =\
e\ f^{a_1 a_2 a_3}\ \k^+_1 \k^+_2\ \dtt{\k}_1\sdot\dtt{\k}_2 \quad.
\eeq


\section{Target Space Actions}

Target space perturbations are induced by source terms in the 
worldsheet action of the form
\beq
\int\!d^2\x\; V^r(X(\x))\ \f_r(X(\x))
\eeq
coupling vertex operators $V^r$ to target space excitations~$\f_r(X)$,
where $X$ stands for generic worldsheet fields.
Hence, there is a one-to-one correspondence between physical string states
$\ket{r;k}$ and spacetime fields $\f_r(x)$.
The on-shell dynamics of the $\f_r$ is determined by tree-level
string scattering amplitudes, $\VEV{V^r\ V^s\ \ldots}$, from which the 
coupling coefficients in the effective target space action can be obtained.
In particular, string three-point functions afford immediate determination
of the cubic terms of the effective target space action thus:
\beq
\int\!d^4k_1\ d^4k_2\ d^4k_3\;\VEV{V^r V^s V^t}(k_i)\
\tilde\f_r(k_1)\ \tilde\f_s(k_2)\ \tilde\f_t(k_3)\ \d(k_1{+}k_2{+}k_3)
\eeq
(here written in momentum space). Further, since the tree-level $N{=}2$ 
string amplitudes vanish for more than three external legs~\cite{hippmann},
the contribution of iterated cubic vertices to $(n{>}3)$-point functions 
must cancel either automatically or in virtue of any necessary further 
terms in the effective target space action. For the Leznov action, it has 
been checked for $n{\le}6$, that such higher-order terms are not 
necessary~\cite{parkes}. Though a general proof does not seem to exist, we 
shall assume that our cubic target space actions are (tree-level) exact.

Taking seriously the higher-spin states of the previous section,
we associate
\beql
\ket{-2{-}2j; (\a_1\a_2\ldots\a_{2j}),\ k} \qquad &\Leftrightarrow &\qquad
\f_{(\a_1\a_2\ldots\a_{2j})}(x) \zeile
\ket{-2{-}2j; e, \q, k} \qquad &\Leftrightarrow &\qquad
v^{\a_1} v^{\a_2} \ldots v^{\a_{2j}}\ \f_{(\a_1\a_2\ldots\a_{2j})}(x)\
\equiv\ \f_{(j)}(x)
\eeql
for $j{\geq}0$ and analogously for $j{<}0$, yielding
a spectrum of target space fields taking values in the Lie algebra 
of the (Chan-Paton) gauge group. Choosing $\q{=}0$, we associate to these
fields their highest $GL(1,\bR)$-boost eigenstates thus:
\beq
\begin{array}{l|ccccccc}
\;\p\quad\;&\quad\cdots\quad&
\quad-4\quad&\quad-3\quad&\quad-2\quad&\quad-1\quad&\quad0\quad&
\quad\cdots\quad\\ \hline \f_{(j)} & \cdots &
\f_{(-1)} & \f_{(-\half)} & \f_{(0)} & \f_{(+\half)} & \f_{(+1)} &
\cdots \\[7pt] \f^{\cdots} & \cdots &
e^{-1}\f^{++} & e^{-\half}\f^+ & \f & e^{+\half}\f^- & e^{+1}\f^{--} &
\cdots \end{array} \quad.
\eeq

The effective action for this infinite tower of fields is surprisingly simple:
\beq
S_\infty\ =\ \int\!d^4x\; \Tr\biggl\{
-\fr12\sum_{j\in\bZ/2} \f_{(-j)}\bo\f_{(+j)}\ +\ \fr16\sum_{j_1+j_2+j_3=1}
\f_{(j_1)}\ \Bigl[ \pa^{+\ad}\f_{(j_2)}\ ,\ \pa^+_\ad\f_{(j_3)} 
 \Bigr]\biggr\}  \la{Sinfty}
\eeq
produces all tree-level string amplitudes correctly.
Interestingly, this action can be truncated to a finite number of fields
in three ways.
First, we may restrict ourselves to $|j|=1$, which is closed under the
interactions. Indeed,
\beq
S_2\ =\ \int\!d^4x\; \Tr\biggl\{
-\f^{++}\bo\f^{--}\ +\  \fr{e}{2}\ \f^{++}\
\Bigl[ \pa^{+\ad}\f^{--}\ ,\ \pa^+_\ad\f^{--} \Bigr]\biggr\}
\eeq
is the two-field action of Chalmers and Siegel~\cite{CS}
for self-dual Yang-Mills in the Leznov gauge~\cite{leznov}.
Concretely, the Leznov field $\f^{--}$ coming from $\p{=}0$
and a multiplier field $\f^{++}$ from $\p{=}-4$ interact via
a $(j_1,j_2,j_3)=(-1,+1,+1)$ vertex~\cite{LS}.
Second, adding $\f{=}\f_{(0)}$ from $\p{=}-2$ yields a three-field action
$S_3$ containing a $(0,0,+1)$ coupling as well.
Third, allowing also the fields $\f_{(\pm\half)}$ yields a
five-field action (boosting $e\to1$)
\beq\begin{array}{rl}
S_5\ =\ {\displaystyle{\int}}\!d^4x\; \Tr\biggl\{ &
          \fr12 \pa^{+\da} \f \pa^-_\da \f\
                    +\  \pa^{+\da} \f^{+} \pa^-_\da \f^{-}\
                    +\  \pa^{+\da} \f^{++} \pa^-_\da \f^{--}\
          \\[5pt]  &
                 +\ \half \f [ \pa^{+\da} \f^{-} , \pa^+_\da \f^{-} ]\
                 +\ \half \f^{--} [ \pa^{+\da} \f , \pa^+_\da \f ]
	  \\[5pt]  &
                 +\     \f^{--} [ \pa^{+\da} \f^+ , \pa^+_\da \f^- ]\
                 +\ \half \f^{++} [ \pa^{+\da} \f^{--} , \pa^+_\da \f^{--} ]\
          \biggr\}  \la{S5}
\end{array}\eeq
for the range $j=-1,\ldots,+1$.
Note that the fields $ \f^{++}$ and $\f^{+}$ effectively play the
r\^ole of (propagating) Lagrange multipliers for the fields
$ \f^{--}$ and $\f^{-}$ respectively.
Although the above actions merely serve to generate the
(classical) background equations of motion,
it is remarkable that the theories based on
$S_2$, $S_3$, and $S_5$ are {\it one-loop exact\/};
their Feynman rules do not support higher-loop diagrams.
Any attempt to include further fields beyond $|j|\leq1$ requires 
the infinite set and no longer forbids two-loop diagrams.

We have seen that the physical states in different pictures
are connected by acting with $\tPCO^+$ (for $\q{=}0$).
Consequently, picture-raising induces a dual operation $Q^+$
on the set of spacetime fields, which lowers the spin by~$\fr12$.
Interestingly, for $j\leq+1$,
\beq
Q^+\ \f_{(j)}\ =\ (3{-}2j)\ \f_{(j-\half)} \la{Qaction}
\eeq
turns out to be a derivation.
By this we mean that the equations of motion for $S_5$,
\beq\begin{array}{lcl}
\pa^{+\da} \pa^-_\da \f^{--}\ &=&\
           \fr12 [ \pa^{+\da} \f^{--} , \pa^+_\da \f^{--} ] \\[5pt]
\pa^{+\da} \pa^-_\da \f^{-}\ &=&\
            [ \pa^{+\da} \f^{-} , \pa^+_\da \f^{--} ] \\[5pt]
\pa^{+\da} \pa^-_\da \f\ &=&\
            [ \pa^{+\da} \f , \pa^+_\da \f^{--} ]\ +\
            \fr12 [ \pa^{+\da} \f^{-} , \pa^+_\da \f^{-} ] \\[5pt]
\pa^{+\da} \pa^-_\da \f^{+}\ &=&\
            [ \pa^{+\da} \f^{+} , \pa^+_\da \f^{--} ]\ +\
            [ \pa^{+\da} \f , \pa^+_\da \f^{-} ] \\[5pt]
\pa^{+\da} \pa^-_\da \f^{++}\ &=&\
            [ \pa^{+\da} \f^{++} , \pa^+_\da \f^{--} ]\ +\
            [ \pa^{+\da} \f^{+} , \pa^+_\da \f^{-} ]\ +\
            \fr12 [ \pa^{+\da} \f , \pa^+_\da \f ] \quad,
\end{array}\la{5}\eeq
all follow from the top one by applying
\beq
Q^+:\ \f^{--}\ \longrightarrow\ \f^-\ \longrightarrow\ 2\ \f\
\longrightarrow\ 2\sdot3\ \f^+\ \longrightarrow\ 2\sdot3\sdot4\ \f^{++}
\eeq
with the Leibniz rule. The reason for these curious facts will become clear 
in the following section.

\vfill\break


\section{Self-dual picture album}

\noindent
{\bf \thesection.1\  Extended Poincar\'e algebra} 

\noindent
It is not often appreciated that the Poincar\'e algebra can have not only 
a $\bZ_2$-graded superalgebra extension, but a $\bZ_2$-graded {\it Lie\/} 
algebra extension as well. The \sp\ of signature $(p,q)$ is a 
$\bZ_2$-graded vector space $\ca=\ca_0 + \ca_1$ 
with a superskewsymmetric bilinear map (super commutator) 
$[.,.] : \ca \times \ca \ra \ca$
such that
$[\ca_\a ,\ca_\b] \subset \ca_{\a+\b}$, with $\a,\b\in \bZ_2$.
Here, $\ca_0$ is the Poincar\'e algebra of signature $(p,q)$,
and $\ca_1$ is a spinor representation of $so(p,q)$ with 
$[\ca_1,\ca_1]\subset \bR^{p,q}$. 
One can have a similar structure with the supercommutator replaced by a
standard skewsymmetric Lie bracket. Thus giving elements of $\ca_1$ the 
`wrong' statistics yields an extension of the Poincar\'e algebra which 
remains a Lie algebra. Such Lie algebra extensions, as well as 
superalgebra extensions, of the Poincar\'e algebra have been classified 
for arbitrary dimension and signature only recently~\cite{ac}, where a 
correspondence is established (theorem 6.2) between superalgebra 
extensions of signature $(p,q)$ and Lie algebra extensions of 
signature $(-q,-p)$ mod $(4,\pm4)$.
For the $(2,2)$ case, there clearly exists an even variant of the standard 
\N1 \sp, with the {\it commutator\/} of two Grassmann-even spinorial 
generators squaring to an $\bR^{2,2}$ translation, 
$[Q_\da,Q_\a] = P_{\a\da}$. 
Indeed, a representation  by vector fields may easily be constructed:
\be \arr
Q_\a &\mapsto\  Q_\a\ =\ \der{\h^\a} + \half \eb^\da \der{x^{\a\da}} \\[5pt]
Q_\da &\mapsto\ Q_\da\ =\ \der{\eb^\da} - \half \h^\a \der{x^{\a\da}} \\[5pt]
P_{\a\da} &\mapsto\ [ Q_\da, Q_\a ]\ =\ P_{\a\da}\ =\ \der{ x^{\a\da}} 
\ea\ee
with the $so(2,2)$ transformations being given in the usual fashion. Here 
$\eta,\eb$ are {\it commuting\/} spinors. It is this Lie algebra which is 
relevant for a covariant description of the spectrum of \N2 string states 
described above.                                                          

\noindent
{\bf \thesection.2\  Hyperspace self-duality}

\noindent
Define a {\it multi-picture hyperspace\/} \hM\ with coordinates 
$ \{ x^{\a\da}, \eb^{\da}, \eta^\a \}$, where
$\{ \eb^{\da}, \eta^\a \} $ are {\it commuting\/} spinorial coordinates
and 
$x^{\a\da}$ are standard coordinates on $\bR^{2,2}$, on which the 
{\it component fields\/} depend.  The Lie algebra described in (\thesection.1)
clearly acts covariantly on this hyperspace.
A {\it \sd\ hyperconnection\/} is subject to the following constraints 
(c.f. \cite{jac,N1,matr,allN})
\bea
 & [\nab_{\da},  \nab_{ \db} ] &=\  \   0 \la{sda}\\[5pt]
 & [\nab_{\da}, \nab_{\b \db}  ]  &=\  \   0 \\[5pt]
 & [\nab_\db ,   \nab_{\a} ] &=\  \     \nab_{\a \db} \\[5pt]
 & [\nab_\a , \nab_\b ] &=\ \ \e_{\a\b}\ \widehat F  \  \la{F} \\[5pt]
 & [ \nab_\a , \nab_{\b \db}  ]  &=\ \ \e_{\a\b}\ \widehat F_{\db}\   \\[5pt]
 & [\nab_{\a\da}, \nab_{\b \db}  ]  &=\ \ \e_{\a\b}\ \widehat F_{\da\db} 
\la{sdf} \quad.
\eea
The first three conditions allow the choice of a {\it chiral basis\/}
in which the covariant derivatives take the form
\bea
 &\nab_\da &=\  \  \pa_\da\  =\ \der{\eb^\da} \\[5pt] 
 &\nab_{\a} &=\  \   \cd_{\a}\ +\ \eb^\da \cd_{\a \da} \\[5pt]
 &\nab_{\a \da}  &=\  \   \cd_{\a \da} \quad, \eea
where $(\cd_{\a}, \cd_{\a \da})$ are gauge-covariant derivatives in the
{\it chiral subspace\/}, \sM, independent of the $\eb^\da$ coordinates. 
In this basis the single constraint \gl{F}
encapsulates the content of all the other constraints. 
The spinorial component of the gauge potential  
\be \widehat A_{\a}(x,\eb,\eta)\ =\
  A_{\a}(x,\eta)\ +\  \eb^\da A_{\a \da}(x,\eta)\ \ee 
describes the entire \sd\ multi-picture hypermultiplet in the form of the
curvature component $ \widehat F $, which has a quadratic
$\eb$-expansion in terms 
of chiral hyperfields of the form
\be \widehat F(x,\eb,\eta)\ =\ 
 F(x,\eta)\ +\ 2 \eb^\da  F_\da (x,\eta)\
 +\ \eb^\da \eb^\db F_{\da\db}(x,\eta) \quad.\la{hatf}\ee
The $\eta$-expansion of $F$ yields an infinite tower of higher
spin fields $\c_\a, g_{\a\b},$ $\psi_{\a\b\g},$ $C_{\a\b\g\d}, \dots $ etc.

\noindent
{\bf \thesection.3\  Extended self-dual component multiplet}

\noindent
Gauge-covariant derivatives in the chiral hyperspace take the form
\be         
\cd_{\a}\ =\ \pa_{\a}\ +\ A_{\a}  \qquad,\qquad
\cd_{\a \da}\ =\ \pa_{\a \da}\ +\ A_{\a\da} \quad,\ee
where the partial derivatives $\ \pa_{\a} \equiv \der{\eta^{\a}},\quad
\pa_{\a \da} \equiv \der{x^{\a \da}}\ $
provide a holonomic basis for the tangent space. 
The components of the gauge connection $( A_{\a}, A_{\a\da})$
take values in the Lie algebra of the gauge group, their transformations
being parametrised by Lie algebra-valued sections on \sM\  
\bea
 \d A_{\a}\ &=&\ - \pa_{\a}\tau(x,\eta)\ -\ [A_{\a},\tau(x,\eta) ]
 \la{sgt}\\[5pt]
 \d A_{\a\db}\ &=&\ - \pa_{\a\db}\tau(x,\eta)\ -\ [A_{\a\db},\tau(x,\eta)]
\quad.\la{trsf}\eea
On \sM, the \sdy\ conditions take the form of the following
curvature constraints
\bea [\cd_{(\a},  \cd_{\b)} ]\ &=&\  0 \la{css}\\[5pt]
 [\cd_{(\a}, \cd_{\b) \db}  ]\  &=&\  0  \la{csv}\\[5pt]
 [\cd_{(\a \da}, \cd_{\b) \db }  ]\  &=&\  0 \quad,\la{cvv}
 \eea
or equivalently
\be\arr
     [\cd_{\a},  \cd_{\b}]\ &=&\ \e_{\a\b} F \\[5pt]
     [\cd_{\a}, \cd_{\b \db}]\ &=&\  \e_{\a\b} F_{ \db}  \\[5pt]
     [\cd_{\a \da}, \cd_{\b \db}]\ &=&\  \e_{\a\b} F_{\da\db} \quad.\la{cvv1}
\ea\ee
Here, $F_{\da\db}=F_{\da\db}(x,\eta)$ is symmetric and has the corresponding 
$\bR^{2,2}$ \ym\ field-strength $F_{\da\db}(x)$ as its leading component in 
an $\eta$--expansion. Henceforth all fields are chiral hyperfields, 
depending on both $x^{\a\da}$ and $\eta^{\a}$.

The non-zero curvature components are not independent; they are related
by super-Jacobi identities. Firstly, the dimension $-3$ \ji\ implies, in
virtue of the constraint \gl{cvv}, the \ym\ equation for the 
{\it hyperfield\/} $F_{\da\db}$,
\be  \cd_\a^{\;\da} F_{\da\db}\ =\ 0 \quad.\la{ym}\ee
Next, in virtue of the constraints \gl{csv} and \gl{cvv}, 
the dimension $-2\half$ Jacobi identity yields 
a dynamical equation for
the dimension $-{3\over 2}$ curvature, 
\be \cd_\a^{\;\da} F_{\da}\ =\ 0 \quad.\la{el}\ee
Finally, the \dim{-2} \ji\ says that
\be \cd_{\a\da} F\ =\ \cd_{\a} F_{\da} \quad.\la{bw1}\ee
We therefore obtain the equation of motion 
\be \square F\ =\  [ F^\da, F_{\da} ] \quad,\la{ew}\ee 
where the covariant d'Alembertian is defined by
$\square \equiv \half\cd^{\a\db}\cd_{\a\db}$. 

Repeated application of $\cd_\a$ on $F$ successively yields an
infinite tower of higher spin fields 
\be \c_\a = \cd_\a F \ ,\quad g_{\a\b} = \cd_{(\a} \cd_{\b)} F \ ,\quad 
\psi_{\a\b\g} = \cd_{(\a} \cd_\b \cd_{\g)}\ F \ ,\quad
C_{\a\b\g\d} =\cd_{(\a} \cd_\b \cd_\g \cd_{\d)}\ F \  , \dots \la{turm} 
\ee 
all having bosonic statistics.
First-order equations of motion for all these fields may be obtained on 
action of $\cd_\a^{\;\da}$ and use of the constraints \gl{cvv1}.
For instance, $\c_\a$ and  $g_{\a\b}$ satisfy
\bea
\cd^\a_{\;\da} \c_\a   \ &=&\ 3\ [ F_\da , F ] \\[5pt]   
\cd^\a_{\;\da} g_{\a\b}\ &=&\  
     \fr{8}{3}\  [ F_\da , \c_\b ]\ +\ 2\  [\cd_{\b\da}F , F ]\quad.
\la{g}
\eea
The entire tower of fields satisfies such gauge-covariant and interacting 
variants of the first-order Dirac--Fierz equations for zero rest-mass fields 
of arbitrary spin~\cite{d,f,p}
\be 
\cd^{\a_n}_{\;\da} \f_{\a_1 \dots \a_{n}}\ =\
J_{\a_1 \dots \a_{n-1}\da}\quad,\qquad n\ge 2\quad.\la{df}
\ee
The interaction current depends on all lower spin fields and is
covariantly conserved, 
\be  \cd^{\a_1\da}J_{\a_1 \dots \a_{n}\da}\ =\ 0\quad,\la{ccc}\ee
in virtue of lower spin field equations. This provides a sufficient condition
for the consistency of the linear equations \gl{df}. In fact, the 
$\eta$-expansion of equation \gl{g} for the hyperfield $g_{\a\b}$ yields
the leading components of all the higher spin equations. 
Now due to the self-duality of the connection \gl{cvv1}, 
$\cd_\b^{\;\da} \cd_{\;\da}^\a = -\d^\a_\b \square$, the gauge 
covariant d'Alembertian. Therefore, covariant derivation of \gl{df} yields 
wave equations for the entire tower of fields of the form
\be 
\square \f_{\a_1 \dots \a_{n}}\ =\  
-\  \cd_{(\a_n}^{\;\;\;\da} J_{\a_1\dots\a_{n-1})\da}\quad,\qquad n\ge 2\quad.
\la{wave}
\ee
The structure of this system of increasingly higher spin interacting fields
is very similar to the arbitrary $N$ extended supersymmetric 
self-dual system presented in \cite{allN}. In fact, just as in that 
supersymmetric case~\cite{dopl}, the covariantly conserved sources \gl{ccc}
provide an infinite set of local conserved currents for this theory,
\be
j_{\a_1 \dots \a_{n-1}\da}\ =\ J_{\a_1 \dots \a_{n-1}\da}\ 
               -\  [ A_{\;\;\da}^{\a_n}, \f_{\a_1 \dots \a_{n}}]\quad ,
               \qquad n\ge 2\quad,\la{cc}
\ee
satisfying
\be  \pa^{\a_1\da} j_{\a_1 \dots \a_{n-1}\da}\ =\ 0\quad.\la{cons}\ee

It is these conserved currents which provide consistent interactions of 
fields of arbitrary spin with fields of lower spin. 
Consistency of higher spin interactions as a consequence of lower spin
field equations is part of the nested structure characteristic of self-dual
systems~\cite{allN}.
In fact, the absence of conjugation in $\bR^{2,2}$
between dotted and undotted spinor indices weakens the
compatibility conditions which, in Minkowski space, make it
almost impossible to construct consistent
gauge-covariant higher spin field equations.
Moreover, the features of having only the
Yang-Mills coupling constant and only the associated spin one gauge
invariance render inapplicable traditional theorems forbidding
higher spin couplings. Just as in the supersymmetric systems
discussed in~\cite{allN}, all our fields \gl{turm} take values in the
Lie algebra of the gauge group and are linear in the 
(dimensionless) Yang-Mills 
coupling constant, which we absorb into the definition of the fields.  
The vector potential transforms in the usual inhomogeneous 
fashion \gl{trsf}, whereas all other fields transform covariantly under
gauge transformations, 
$\d \f_{\a_1 \dots \a_{n}} = [\tau , \f_{\a_1 \dots \a_{n}} ] $. 
These are the only gauge transformations of these fields;
there are no higher spin gauge invariances. The latter are not
required since all fields apart from the vector potential
transform according to irreducible representations of $SO(2,2)$.
They therefore do not contain any redundant degrees of freedom, which
would have required elimination in virtue of further (higher spin) 
gauge invariances.

\noindent
{\bf \thesection.4\  Prepotential and action}

\noindent
As we have already mentioned, the NSR formulation of the \N2 string requires 
a specific choice of complex coordinates, leading to the breaking \gl{subgp}
of $SO(2,2)$. A convenient field-theoretical tool for describing complex 
structures is that of harmonic spaces~\cite{harm}. 
This is a covariant description of the space of complex structures,
the coset space $SL(2,\bR)/GL(1,\bR)$, given by equivalence classes under 
the parabolic $GL(1,\bR)$ subgroup. The quotient has homogeneous coordinates
which may be organised into spinors $u^{\pm\a}$ satisfying  $u^{+\a}u^-_\a=1$.
These `harmonics' provide a covariant version of the two-parameter 
description \gl{vwahl}  of the space of complex structures. A particular 
choice of complex structure corresponds to choosing specific 
spinors~$u^{\pm\a}$. However, in the harmonic space method,
these spinors are treated as independent variables, and
this coset space is adopted as an auxiliary space with vector fields,
\be
\arr  \pa^{++}\ &=&\ u^{+ \a} \der {u^{- \a}}  \\[5pt]
         \pa^{--}\ &=&\ u^{- \a} \der {u^{+ \a}} \\[5pt]
 \pa^{+-}\ &=&\ u^{+\a} \der {u^{+\a}} - u^{-\a} \der {u^{-\a}}\quad,
\ea\ee
satisfying the $sl(2,\bR)$ algebra. 
Although this coset is non-compact and there certainly exist subtleties, 
we may apply formal rules of harmonic analysis on it (see e.g.~\cite{e}), 
understanding these in the sense of a Wick-rotated version of those applying
to the compact case. The application of harmonic space methods to our
hyperspace self-duality equations follows the treatment of other self-duality
related systems, reviewed for instance in~\cite{guer}.

In analogy to the supersymmetric case \cite{N1} we 
enlarge \hM\ to a \hs\ with coordinates
$\{ x^{\da\pm}, \eb^\da , \eta^{\pm},  u^\pm_\a \} ,$
where $ x^{\da\pm} = u^\pm_\a x^{\a\da},$
and $\eta^{\pm} = u^\pm_\a \eta^{\a} $.
The \hs\  gauge covariant derivatives are given by
\be\arr   
           \cd_\da^\pm \ & = &\ \pa_\da^\pm \ +\ A_\da^\pm   \\[5pt]
            \nab^\pm \ & = &\ \pa^\pm\ +\ A^\pm   \\[5pt]
            \nab_\da\ &=&\  \pa_\da\ +\ A_\da \\[5pt]   
             \cd^{\pm\pm}&\ = &\  \pa^{\pm\pm}\  +\ A^{\pm\pm}\quad , 
\ea\ee
where $[\partial_{\dot\alpha} , \partial^\pm ] = \partial^\pm_{\da}\ $.
The equations \gl{sda}-\gl{sdf} are then equivalent to the following
curvature constraints in \hs:
\be\arr 
  && [\nab_{\da},  \nab_{\db} ] =   0\quad  ,\qquad 
     [\nab_{\da}, \cd^\pm_{\db}  ]  =   0\quad  ,\qquad  
     [\nab_{\db} , \nab^\pm ] =    \cd^\pm_{\db} \\[5pt]
  &&      [\nab^+ , \cd^+_\db ] = 0 \quad  ,\qquad 
          [\nab^- , \cd^-_\db ] = 0 \\[5pt]  
 &&     [\cd^+_\da , \cd^+_\db ] = 0 \quad  ,\qquad
        [\cd^+_{[\da} , \cd^-_{\db]} ] = 0 \quad  ,\qquad
        [\cd^-_\da , \cd^-_\db ] = 0 \la{hsd}\ea\ee
together with the definitions of the non-zero curvatures
\be\arr [\nab^+ , \nab^- ]\ &=&\ F \\[5pt]
        [\nab^+ , \cd^-_\db ]\ &=&\ [\cd^+_\db  , \nab^- ]\ =\ F_\db \\[5pt]
        [\cd^+_\da , \cd^-_\db ]\ &=&\  F_{\da\db} \quad.\la{hcurv}\ea\ee    
An equivalent set of curvature constraints is:        
\be\arr 
 & [\nab_{\da},  \nab_{\db} ] =   0\quad  ,\qquad 
 & [\nab_{\da}, \cd^\pm_{\db}  ]  =   0\quad  ,\qquad 
  [\nab_{\da}, \Dpm ]  =   0 \\[5pt]
 & [\nab_{\db} ,\nab^\pm ] =   \cd^\pm_{\db}\quad  ,\qquad 
  &      [\nab^+ , \cd^+_\db ] = 0\quad  ,\qquad 
        [\cd^+_\da , \cd^+_\db ] = 0   \\[5pt]       
    &    [ \Dmm , \nab^+ ] = -\ \nab^- \quad  ,\qquad
     &   [ \Dmm , \cd^+_\da ] = -\ \cd^-_\da \\[5pt] 
    &    [ \Dmm , \nab^- ] = 0\quad  ,\qquad 
     &   [ \Dmm , \cd^-_\da ] = 0 \quad.\la{hsd1}\ea\ee
The proof of equivalence is immediate in the {\it central frame\/}
defined by $\Dpm = \dpm$, i.e. $A^{\pm\pm} = 0$, 
in which \gl{hsd1} has the partial solution
$\cd^\pm = u^{\pm\a} \cd_{\a}\  ,\  \cd^\pm_\da = u^{\pm\a} \cd_{\a \da }\ $.
\goodbreak
 
The advantage of using \hs\ coordinates is that the existing 
flat subspaces thus become manifest, allowing the choice of a 
{\it Frobenius frame\/} in which $A_\da$, $A^+ $ and $ A^+_\db$ are zero, 
i.e. $\nab_{\da}, \nab^+ $ and $ \cd^+_\db$ are the derivatives 
$\pa_\da, \pa^+ $ and $ \pa^+_\db$, respectively.
In this frame the chiral (i.e. independent of $\eb$) hyperfield 
$\F^{--} = \Amm (x,\eta) $ becomes fundamental,
determining all other fields occuring in the above constraints thus: 
\be\begin{array}{lll}  A^-\ &=\ \pa^+ \F^{--}\quad,\qquad 
        & A^-_\da\  =\ \pa^+_\da \F^{--} \\[5pt]
          F\ &=\  \pa^+ \pa^+ \F^{--}\quad,\qquad 
        & F_\da\  =\ \pa^+_\da \pa^+ \F^{--}\quad,\qquad 
         F_{\da\db}\ =\ \pa^+_\da \pa^+_\db \F^{--} \quad.\la{prepot}\ea\ee
Higher spin fields then arise on iterative application of $\nab^\pm$
according to \gl{turm}. In this frame, most of the constraints in \gl{hsd}
are resolved and the only remaining dynamical equations for $\F^{--}$ are
\be  
\pa^{+\da} \pa^-_\da \F^{--}\  =\ \
           \fr12\ [ \pa^{+\da} \F^{--} , \pa^+_\da \F^{--} ]\quad,\la{dynF}\ee
\be  \pa^+ \pa^-_\da \F^{--}\ -\  \pa^- \pa^+_\da \F^{--}\ =\ \
            [ \pa^+ \F^{--} , \pa^+_\da \F^{--} ]\quad.\la{dynF1}\ee    
These equations are not independent; the former is obtained by acting on the
latter by $\pa^\da$. Moreover, they are also not independent of the
equations for $\F^{--}$ following from the last two constraints in \gl{hsd1}.
Eq. \gl{dynF} is the Euler-Lagrange equation 
for the generalised Leznov functional (c.f. \cite{leznov,dl})
\be\arr
\cl^{----}\ &=&\  
   \Tr\  \pa^\da \left( {1\over 4} \pa^{[-} \F^{--} \pa^{+]}_\da \F^{--} \
               +\ {1\over 6} \F^{--} [ \pa^+ \F^{--} , \pa^+_\da \F^{--} ]    
              \right) \\[5pt]  
   &=&\ \Tr  \left( \half \pa^{+\da} \F^{--} \pa^-_\da \F^{--} \
          +\ {1\over 6} \F^{--} [ \pa^{+\da} \F^{--} , \pa^+_\da \F^{--} ]
                \right)\quad.\la{leznov}\ea\ee    

One advantage of choosing \gl{dynF} and \gl{dynF1} to be the equations 
determining the dynamics is that in this frame the harmonic variables 
$u^\pm$ may be treated as parameters (this explicitly breaks the $SO(2,2)$ 
invariance) and the explicit $u$-dependence may be ignored, treating
$\{x^{\a\da}\} \rightarrow \{x^{\pm\da}\} $ as a fixed choice of complex
structure. As we have seen, this is precisely the choice required for
a comparison with string theory. Now, let us consider explicitly breaking 
the hyperspace-covariance of our theory by considering $\F^{--}$ to be 
independent of $\eta^+$. It may then be Laurent--expanded in powers of 
$\eta^-$ thus
\be  \F^{--}\ =\ \dots + {1\over \eta^-}\f^{---} + \f^{--}  + \eta^- \f^{-} 
           +  (\eta^- )^2 \f  +  (\eta^- )^3 \f^{+}  +  (\eta^- )^4 \f^{++} 
               +  (\eta^- )^5 \f^{+++}  \dots\quad. \la{atlas}\ee
Inserting this expansion into \gl{dynF} yields component equations of 
motion, which in general have infinitely many terms. Remarkably, inserting 
it into \gl{leznov} and picking out the coefficient of $(\eta^- )^4$ 
yields precisely the charge-zero (homogeneous) lagrange functional in 
$S_\infty$ \gl{Sinfty}. So the chiral hyperfield  prepotential 
$\F^{--}(x,\eta^-)$ 
is seen to be a generating function for the entire tower of physical states
of the \N2 string, and the (dual) picture-lowering $Q^+$~\gl{Qaction} 
corresponds to a transformation from the coefficient at one order 
in an $\eta^- $--expansion to the coefficient at the next order.

Denoting the hyperfield having leading (i.e. $\eta^-$-independent) component 
$\f^{\dots}(x)$  by  $\F^{\dots}(x,\eta^-)$, we remark that the leading term 
of the spinorial gauge potential is precisely the $\eta^-$-coefficient,
\be  A^-\ =\  \ \F^-\ =\  \ \pa^+ \F^{--}\quad,\ee
and the coefficient of  $(\eta^- )^2$ is the leading term of the curvature
component $F =\F = \pa^+ \pa^+ \F^{--}$. For $\eta^+$-independent $\F^{--}$, 
\gl{dynF1} takes the form
\be   \pa^-_\da \F^{-}\ =\ 
            [  \F^{-} , \pa^+_\da \F^{--} ]\quad.\la{dynF-}\ee
Acting on both sides by $\pa^{+\da}$ yields the hyperfield version of  
the second equation in \gl{5}. 

Now, if $\F^{--}$ is expandable as a {\it positive\/} power series in 
$\eta^-$, 
\be  \F^{--}\ =\  \f^{--}  + \eta^- \f^{-} 
           +  (\eta^- )^2 \f  +  (\eta^- )^3 \f^{+}  +  (\eta^- )^4 \f^{++} 
               +  (\eta^- )^5 \f^{+++}  \dots\quad,\la{atlas+}\ee
the system of component equations are related by field transformations
implied by \gl{Qaction}, 
which amount to the action of $\pa^+$ on the corresponding hyperfields.

The restricted system of five fields $\f^{--}, \f^- ,$ $\f ,$ $\f^+ ,$ 
$\f^{++} $ satisfies the rather distinguished set of equations \gl{5} 
obtained from the action $S_5$ \gl{S5}. An $SO(2,2)$-covariant 
action for this theory of five fields is given by 
\be
S\ =\ \int d^4x\  \Tr \left( \fr{1}{4} g^{\a\b}F_{\a\b}\  
                        +\  \fr{1}{3} \c^\a \cd_{\a\da} F^\da\
                  	+\  \fr{1}{8} \cd^{\a\da} F  \cd_{\a\da} F\ 
                  	+\  \fr{1}{2} F [ F^\da , F_\da ] \right)\quad,
\la{5field}\ee                     	 
where  $g_{\a\b}$ and $\c_\a$ play the r\^ole of (propagating) Lagrange 
multipliers for $A_{\a\da}$ and $F_\da$ respectively.  This is in fact
very reminiscent of the supersymmetric \N4 action of \cite{siegel}, with 
adjustments made for our different type of extension. 
Under this five-field truncation of the infinite system, only the first
of the conserved currents in \gl{cc} survives, namely, $j_{\a\da}$,
the source current for the spin-one field $g_{\a\b}$ \gl{g}, which is 
the Noether current corresponding to global gauge invariance of the action
\gl{5field}.  
                       

\section{Concluding remarks}

We have seen that a novel extension of self-dual Yang-Mills theory to 
a hyperspace with Grassmann-even spinorial auxiliary coordinates affords 
a covariant description of the physical degrees of freedom of the \N2 open 
string. It yields, moreover, a compact description of the infinite number of 
massless string degrees of freedom  in terms of a scalar hyperspace 
prepotential, for which the generalised Leznov functional \gl{leznov} yields
the action $S_\infty$ \gl{Sinfty} describing the tree-level \N2 string 
amplitudes.
The infinitely large multiplet of interacting massless higher spin fields is 
analogous to the \N{\infty} supersymmetric self-dual multiplet presented 
in~\cite{allN}. In fact the multiplets described by the three consistent 
truncations of $S_\infty$, namely, $S_2, S_3$ and $S_5$, are remarkably 
reminiscent of the supersymmetric \N{1,2,4} self-dual multiplets, 
respectively \cite{N1,matr,siegel}. There appears to exist a 
correspondence between these pairs of theories.    

Our infinite extention of the self-dual Yang-Mills system is amenable
to solution by a twistor-type transform. In fact both the Ward splitting
method and the ADHM construction yield themselves to modifications to 
accommodate our extension. Moreover, twistor theory makes intimate use of the
sequence of zero-mass field equations of spin ${m\over 2}$ ($m{\ge}0$) 
in a self-dual Yang-Mills background and of the associated space of solutions 
to the d'Alembert equation. These are just the sets of equations \gl{df}
and \gl{wave}, respectively, with the interaction currents 
$J_{\a_1 \dots \a_{n-1}\da}$ set to zero. There is thus a tantalising 
similarity between the BRST-cohomological analysis yielding the tower of
\N2 string states and the cohomological description of certain spaces used
in twistor theory (see, for instance, \cite{coho}). We expect the 
interrelationship to be a fruitful direction for future research.
The theories of \N2 closed as well as \N{(2,1)} heterotic strings are also 
intimately related to self-dual geometry, and we expect our covariant 
description to generalise to both these cases.

\noindent
{\bf Acknowledgments} 

\noindent
We have benefitted from discussions with D.V.~Alekseevsky, V.~Cort\'es,
K.~J\"unemann and M.A.~Vasiliev. 
C.D. thanks the Institut f\"ur Theoretische Physik der Universit\"at Hannover
for generous hospitality.

\noindent
{\bf Note Added } 

\noindent
The tower of higher spin fields (4.24), evaluated at $\eta^\a =0$, 
correspond, in an appropriate gauge, to coefficients of $F(x,\eta)$ in an 
$\eta^\a$-expansion. The hyperspace field $F(x,\eta)$ can therefore be 
thought of as an $\bR^{2,2}$ field, taking values in the infinite dimensional 
algebra spanned by polynomials of $\eta^\a$. Such algebras have been 
investigated by M. Vasiliev. In particular, he showed that consistent 
higher-spin free-field equations arise as components of zero-curvature 
conditions for connections taking values in such algebras. A description 
and further references can be found in his review article~\cite{v}. 

\end{document}